# Unravelling of the chemistry and the performance in the oxygen reduction reaction of carbon nitride-supported bimetallic electrocatalysts through X-ray photoelectron spectroscopy

Stefano Diodati,[a] Enrico Negro,[a,c] Keti Vezzù,[d] Vito Di Noto[d,b*] and Silvia Gross[a,b]*

[a]Dipartimento di Scienze Chimiche, Università degli Studi di Padova, Via Marzolo 1, 35131 Padova, Italy

[b]Istituto per l'Energetica e le Interfasi, IENI-CNR, Dipartimento di Scienze Chimiche, Università degli Studi di Padova, via Marzolo 1, 35131- Padova and INSTM, UdR di Padova, Italy

[c] Centro Studi di Economia e Tecnica dell'Energia «Giorgio Levi Cases», 35131 Padova (PD) Italy.

[d] Section of Chemistry for Technology, Department of Industrial Engineering, Università degli Studi di Padova in Department of Chemical Sciences, Via Marzolo 1, 35131 Padova (PD) Italy.

stefano.diodati@unipd.it

enrico.negro@unipd.it

keti.vezzu@unipd.it

vito.dinoto@unipd.it

silvia.gross@unipd.it

Five bimetallic electrocatalysts (ECs) including a carbon nitride (CN) support are synthesised through the pyrolysis of a solid precursor obtained through sol-gel and gel-plastic processes. The resulting ECs are characterised through ICP-AES (Inductively Coupled Plasma-Atomic Emission Spectroscopy) and XPS (X-ray Photoelectron spectroscopy); their performance and reaction mechanism in the oxygen reduction reaction (ORR) are evaluated with the CV-TF-RRDE method (Cyclic Voltammetry Thin-Film Rotating Ring-Disk Electrode). Special attention is given to XPS results with the aim to carry out a thorough investigation of the surface concentration and the chemical environments of the different elements, as well as providing information on the structure of the metal components of the ECs and their interactions with the carbon nitride support. The correlation of the results obtained from the chemical analyses, XPS and the electrochemical studies allows to improve the fundamental understanding of the factors controlling the ORR kinetics and reaction pathway in bimetallic CN-supported ECs.

## 1. Introduction

Polymer electrolyte membrane fuel cells (PEMFCs) are a family of energy-conversion devices characterised by a very high efficiency {Vielstich, 2003 #933}. This feature, together with remarkable energy and power densities, make PEMFCs very attractive as energy sources for several applications ranging from light-duty vehicles to portable electronics {Fiori, 2015 #1049}{Zhang, 2012 #1050}. One of the major bottlenecks in PEMFC operation is the sluggishness of the oxygen reduction reaction (ORR), which introduces large overpotentials and thus degrades significantly the energy conversion efficiency of the entire power plant {Katsounaros, 2014 #1052;Zheng, 2012 #1051}. In addition, at the low temperatures typical of PEMFC operation (T < 130°C), the only viable cathode electrocatalysts (ECs) require a significant loading of platinum group metals (PGMs) {Guo, 2013 #1053}{Gasteiger, 2005 #1054}, whose abundance in Earth's crust is very limited {Moss, 2013 #1055}. Thus, the development of enhanced ORR ECs is one of the most important goals of PEMFC research {Gasteiger, 2005 #1054}. One way to enhance the performance of ORR ECs is to use active sites based on a PGM (*e.g.* Pt or Pd) in conjunction with a second element M, generally a first-row transition metal such as Fe, Co, Ni, Cr or others {Guo, 2013 #1053}{Gasteiger, 2005 #1054}. There is considerable debate on the origins of the observed improved performance in PGM-M active sites, which was interpreted in terms of electronic effects originating from the alterations in the crystal structure of the PGM induced by the second element {Guo, 2013 #1053}{Wu, 2013 #1056}. It was also proposed that the enhanced ORR performance might arise from bifunctional effects {Di Noto, 2010 #1057}{Di Noto, 2015 #1058}. In detail the M element, which is usually a good Lewis acid, is assumed to improve the ORR kinetics as it enhances the protonation of ORR intermediates. This facilitates the desorption of the reaction products and enhances the turnover frequency on the active sites {Di Noto, 2010 #1057}{Di Noto, 2015 #1058}. Our research group proposed an innovative procedure for the preparation of ORR ECs {Di Noto, 2010 #1057}{Di Noto, 2015 #1058}{di Noto, 2007 #946}. Several materials were synthesised and extensively characterised, yielding a promising ORR activity {Di Noto, 2010 #1057}{Di Noto, 2015 #1058}. However, until now no effort was reported to identify systematic trends between the atomic ratios, chemical nature and environment of the metal species in the ECs and the observed electrochemical performance. In this work, new and accurate electrochemical measurements are collected with the Cyclic Voltammetry Thin-Film Rotating Ring-Disk Electrode (CV-TF-RRDE) method {Vielstich, 2003 #933}, to determine the performance and reaction mechanism in the ORR. The prepared ECs also undergo a thorough characterisation by X-ray photoelectron spectroscopy (XPS). In particular, the XPS investigation is aimed at determining the oxidation state(s) of the different metals, their interaction with the carbon nitride matrix support, their surface atomic ratios (which are typically different from those determined in the bulk) and their chemical environment as well as correlating them with the observed electrochemical features.

## 2. Experimental

### 2.1 Reagents

Potassium tetrachloroplatinate (II), 99.9%, potassium tetrachloroaurate (III), 98%, rhodium (III) trichloride, hydrate (Rh assay: 38%), potassium hexachloroiridate (III), hydrate (Ir assay: 33.4%) and potassium tetracyanoplatinate (II), hydrate (Pt assay: 45.34%) are obtained from ABCR GmbH. Potassium tetracyanonickelate (II) hydrate, purum is acquired by Fluka. D(+)-sucrose, biochemical grade is an Acros reagent. EC-20 is received from ElectroChem, Inc. (nominal Pt loading: 20%) and is used as a reference EC; in the following text, EC-20 is labelled "Pt/C reference". Vulcan XC-72R carbon black is provided as a courtesy by Carbocrom s.r.l. and washed with $H_2O_2$ 10 vol.% prior to use. All the other reagents are used as received. Ultrapure Milli-Q water (Millipore) is used throughout the synthesis of the materials.

## 2.2 Preparation of the materials

The protocol used to synthesise the materials presented in this work is described in detail elsewhere {Di Noto, 2010 #939;di Noto, 2007 #946}{Di Noto, 2009 #947}{Di Noto, 2007 #948}{Di Noto, 2010 #945}{Di Noto, 2007 #944}{Di Noto, 2007 #943}{Di Noto, 2007 #942}{Di Noto, 2007 #940}. The first step consists in the preparation of a solid precursor, according to the following general procedure. An aliquot of a coordination compound including a metal atom A bearing chloride ligands (*e.g.*, $K_2PtCl_4$, $KAuCl_4$, $RhCl_3{\cdot}xH_2O$ or $K_3IrCl_6{\cdot}xH_2O$) is dissolved into water, yielding a clear solution (Solution A). An aliquot of a coordination compound including a metal atom B bearing cyano ligands (e.g., $K_2Pt(CN)_4{\cdot}xH_2O$ or $K_2Ni(CN)_4{\cdot}xH_2O$) is dissolved into water, yielding a second clear solution (Solution B). An aliquot of sucrose is dissolved into water, yielding a further clear solution (Solution C). The latter is then divided equally between Solution A and Solution B, yielding the two clear solutions (A+C) and (B+C). (A+C) and (B+C) are then mixed together, yielding solution (A+B+C), which is stirred for 15 minutes and then allowed to rest overnight. The amounts of A and B reactants, the amount of sucrose and the nominal molar ratio between A and B are reported in Table S1 in SI. The solid precursor is obtained after removing the liquid fraction from the resulting product with a rotovapour at T = 65°C.

Each solid precursor is dried at 120 °C for 16 h and pyrolysed at 300 °C for 2 h, yielding a deep brown, coarse material (D). The whole thermal procedure is carried out under a dynamic vacuum of $10^{-3}$ bar. (D) is ball-milled for 3 h and thermally treated at 600 °C under a dynamic vacuum of $10^{-3}$ bar for 2 h. The resulting product consists in a black powder (E) which is ball-milled for 2 h, washed three times with ultrapure water, treated with $H_2O_2$,10 vol.% and eventually dried under an IR lamp, giving rise to the final EC product. The ECs are labelled in accordance with the nomenclature proposed elsewhere {Di Noto, 2010 #1057}{Di Noto, 2015 #1058}; a total of five ECs are obtained, as follows: PtIr-$CN_l$ 600, $PtNi_{1.9}$-$CN_l$ 600, PtRh-$CN_l$ 600, AuNi-$CN_l$ 600 and IrNi-$CN_l$ 600.

In AuNi-$CN_l$ 600 and IrNi-$CN_l$ 600 the *"active metal"* is Au and Ir, respectively. In all the other ECs and in the Pt/C reference, the *"active metal"* is Pt {Di Noto, 2015 #1058}. For all the ECs, the other metal species is referred to as the *"co-catalyst"*.

## 2.3 Instruments and Methods

Each ball-milling step lasts 30 minutes and is carried out in a Retsch PM 100 planetary mill mounting an agate jar. The metal composition of the ECs is determined by ICP-AES using the method of standard additions. The emission lines are: λ(Pt) = 214.423 nm, λ(Rh) = 343.489 nm, λ(Ir) = 224.268 nm, λ(Au) = 267.595 nm, λ(Ni) = 352.454 nm, λ(K) = 766.490 nm. A Spectroflame Modula sequential and simultaneous ICP-AES spectrometer equipped with a capillary cross-flow nebuliser is used (Spectro Analytical, Kleve, Germany). Analytical determinations are carried out using a plasma power of 1.2 kW, a radio frequency generator of 27.12 MHz, and an argon gas flow with nebuliser, auxiliary, and coolant set at 1, 0.5, and 14 L·min$^{-1}$, respectively. The details on the procedure for the digestion of the samples and the elemental analyses are reported elsewhere {Di Noto, 2007 #948}.

## 2.4 Electrode Preparation and Electrochemical Measurements

Porous electrodes are prepared according to the procedure described by Di Noto *et al.* {Di Noto, 2007 #942}. Each EC is diluted with XC-72R in a 1:1 weight ratio; the resulting dispersion is ground in an agate mortar until a homogeneous black powder is obtained. EC inks are prepared by adding suitable amounts of milli-Q water and commercial Nafion solution (Alfa Aesar, 5% weight) to these dispersions. 15 μL of each ink are then transferred to the top of a freshly-polished glassy carbon electrode tip with an active diameter of 5 mm, yielding an overall loading of 15 $\mu g \cdot cm^{-2}$ of the active metal. Water is then removed from the EC ink by evaporation under an IR lamp. The same procedure is applied to prepare the electrode to test the Pt/C reference, with the difference that no XC-72R is added. The electrode tip also mounts a coaxial platinum ring electrode characterised by a collection efficiency N equal to 38%. The tip is mounted on a Model 636 rotating ring-disk electrode system produced by Pine Research Instrumentation, connected to a multi-channel VSP potentiostat/galvanostat produced by BioLogic. The resulting system is used as working electrode in order to test the current densities of the ECs in the ORR. The cell is filled with a 0.1 M $HClO_4$ solution and the electrochemical measurements are carried out at 60°C. A platinum counter-electrode is adopted. An $Hg/HgSO_4/K_2SO_{4(sat.)}$ reference electrode is placed in a separate compartment and connected to the main electrochemical cell by a salt bridge. The ECs are activated by cycling the working electrode at 50 $mV \cdot s^{-1}$ between 0.04 and 1.14 V *vs.* RHE (Reversible Hydrogen Electrode) until the voltammogramms are stable. High-purity oxygen (Air Liquide) is used to saturate the electrochemical cell to investigate the performance and the reaction mechanism of the proposed ECs in the ORR. Cyclic voltammogramms are collected at 5 $mV \cdot s^{-1}$ from 0.04 to 1.14 V *vs.* RHE as the electrode tip is rotated at 1600 rpm; the platinum ring is polarised at 1.2 V *vs.* RHE. Data reproducibility is tested by preparing three different electrodes with each electrode ink; afterwards, each electrode is used to collect measurements using the same protocol.

**2.5 X-Ray photoelectron spectroscopy analysis**

The composition of the ECs is investigated by XPS. XP spectra are run on a Perkin-Elmer ϕ5600ci spectrometer using Al monochromatised radiation (1486.6 eV) working at 350 W. The working pressure is $< 5 \cdot 10^{-8}$ Pa. The spectrometer is calibrated by assuming the binding energy (BE) of the $Au4f_{7/2}$ line at 83.9 eV with respect to the Fermi level. The standard deviation for the BE values is 0.15 eV. The reported BE are corrected for the charging effects, assigning, in the outer layers where contamination carbon is still present, to the C1s line of carbon the BE value of 284.6 eV {Briggs, 1990 #459}{Moulder, 1992 #460}. Survey scans (187.85 pass energy, 1 eV/step, 25 ms per step) are obtained in the 0-1300 eV range. Detailed scans (58.7 eV pass energy, 0.05-0.1 eV/step, 100-200 ms per step) are recorded for the O1s, C1s, N1s, Cl2p, S2p for all the samples, whereas the other region of the elements of interest (Au, Ir, Pt, Pd, Ni, Rh) are different among the samples.

The atomic composition was evaluated using sensitivity factors supplied by Perkin-Elmer after the subtraction of a Shirley type background {Shirley, 1972 #462}{Moulder, 1992 #460}. The assignment of the peaks is carried out by using the values reported in the literature {Moulder, 1992 #460} and in the NIST XPS Database {,  #458}

For a detailed analysis, the core-level lines obtained by XPS are fitted by the freeware program XPS Peak 4.1 by subtracting a Shirley background and by using Gaussian-Lorentzian contributions. During the fitting procedure. Elemental analyses are carried out by using the home-made program XPS HITS Version 4.7 (2008), adopting the corresponding photoioinisation cross-sections taken from the literature {Moulder, 1992 #460}.

## 3. Results and discussion

### 3.1 Bulk and surface composition of the ECs

Microanalysis and inductively-coupled plasma atomic emission spectroscopy (ICP-AES) measurements allow to determine the bulk chemical composition of the ECs. Results are summarised in Table 1.

| Material | Wt% | | | | | | | | | Formula |
|---|---|---|---|---|---|---|---|---|---|---|
| | Pt[a] | Au[a] | Ir[a] | Rh[a] | Ni[a] | K[a] | C[b] | N[b] | H[b] | |
| *AuNi-CN$_l$ 600* | - | 21.13 | - | - | 4.82 | 0.59 | 49.33 | 1.28 | 1.29 | $K_{0.14}[AuNi_{0.77}C_{38.29}N_{0.85}H_{11.93}]$ |
| *IrNi-CN$_l$ 600* | - | - | 10.23 | - | 2.70 | - | 42.61 | 2.22 | 1.13 | $IrNi_{1.84}C_{66.64}N_{2.98}H_{21.06}$ |
| *PtIr-CN$_l$ 600* | 11.1 | - | 12.1 | - | - | 1.50 | 53.93 | 2.12 | 1.00 | $K_{0.67}[PtIr_{1.1}C_{79}H_{17}N_{2.7}]$ |
| *PtNi$_{1.9}$-CN$_l$ 600* | 16.9 | - | - | - | 9.72 | 2.12 | 50.28 | 2.19 | 0.75 | $K_{0.6}[PtNi_{1.9}C_{48}N_{1.8}H_{8.6}]$ |
| *PtRh-CN$_l$ 600* | 15.21 | - | - | 6.84 | - | 0.85 | 60.17 | 1.72 | - | $K_{0.28}[PtRh_{0.85}C_{64}N_{1.6}]$ |
| *Pt/C reference[c]* | 20 | - | - | - | - | - | 80 | - | - | $[PtC_{65}]$ |

[a] Determined by ICP-AES spectroscopy. *b* Determined by microanalysis. *c* Reference material supplied by Electrochem Inc. Nominal values.

**Table 1** Chemical composition of the ECs and of the Pt/C reference.

It is noticed that all the ECs are characterised by a N wt% ranging between *ca.* 1.3 and 2.2. This evidence shows that the incorporation of nitrogen in the chemical structure of the materials is successful, giving so rise to a carbon nitride (CN) support with a relatively low concentration of nitrogen {Di Noto, 2007 #948}{Negro, 2008 #941}. Most ECs are characterised by a small concentration of hydrogen atoms; this indicates that the graphitisation process of the support is not complete. ECs for application in PEMFCs should feature a high electrical conductivity to prevent the introduction of ohmic drops lowering the overall performance of the final device {Di Noto, 2015 #1058}. The literature {Negro, 2008 #941} shows that other CN-supported ECs prepared according to the same protocol followed in this paper and characterised by a similar concentration of N and H display a remarkable performance in operative conditions at the cathode of a PEMFC {Di Noto, 2015 #1058}. For this reason, it is expected that the ECs presented in this paper feature a sufficiently high electron conductivity for their intended application at PEMFC electrodes. The support of the ECs is also expected to stabilise the metal species in *"coordination nests"* obtained from C and N atoms of the CN matrix, as observed in other similar systems

{Di Noto, 2014 #1060}{Negro, 2014 #1059}. Potassium is detected in the chemical composition of most of the ECs presented in this paper. This trend is typical of this family of materials {Negro, 2014 #1059}{Di Noto, 2015 #1058} and is attributed to the presence of surviving anionic metal coordination complexes in the materials arising from the incomplete reduction of the metal ions during the graphitisation process. Thus, $K^+$ ions are necessary to counterbalance the electrical charge of coordination metal complexes and consequently are not removed during the washing process {Di Noto, 2007 #948}. It is observed that AuNi-$CN_l$ 600 and IrNi-$CN_l$ 600 evidence the lowest concentration of potassium among the investigated materials. This is an indication that Au and Ir are easily reduced to the (0) oxidation state during the pyrolysis process of the precursors, as further confirmed by XPS analyses (*vide infra*). In the following discussion, and in accordance with Section 2.2., the *"main"* metal of each EC is indicated as *"M1"*. On the other hand, the *"co-catalyst"* is indicated as *"M2"*. It should be noted that for all the ECs the stoichiometry of the M1 and M2 metals is quite similar to that of the starting reagents; this is a further proof that the proposed preparation protocol allows for the synthesis of ECs with a well-controlled chemical composition {Di Noto, 2015 #1058}. The only exception is IrNi-$CN_l$ 600, where an unexpected decrease in iridium concentration is detected. The chemical reactions taking place in the preparation of the precursor of this family of ECs are well-documented {Di Noto, 2007 #948}{Di Noto, 2007 #942}{Di Noto, 1997 #937}{Di Noto, 2004 #936}{Di Noto, 2000 #935}{Di Noto, 2003 #934}, and include the establishment of cyano bridges between the two different metal species present in the reaction mixture. During this process, the nitrogen atom terminating a -C≡N group displaces one of the chloride ligands of the other complex. Nevertheless, a weak complexation process of Ir atoms by the cyanide group nitrogen is likely responsible for a low concentration of Ni-C≡N→Ir in the precursor of IrNi-$CN_l$ 600. Consequently, a significant fraction of the $K_3IrCl_6$ complexes originally present in the reaction mixture is removed during the washing process.

XPS analysis is adopted to investigate the surface composition and the chemical environments of the species present on the synthesised materials. In particular, these analyses are carried out with the objective to identify the surface oxidation states of the M1 and M2 metals, as well as to obtain information concerning the nature of metal species on the surface (pristine metal, hydroxides etc.). The correlation between the surface concentration and the chemical environment also allows to gain insight one the nature of the obtained bimetallic systems (alloys, "*core-shell*" particles *etc*...). An additional aim of this investigation is to elucidate the chemical interactions between the matrix and the metal species. In a first step, a survey spectrum is collected from each sample to identify the main elements present. Subsequently, high-resolution multiplex spectra are collected focusing on the regions of interest for both: (i) the main elements (carbon, oxygen, nitrogen and the A and B metals); and (ii) additional elements evidenced in the survey spectrum and arising from residues of the precursor. On this basis, the quantitative evaluation (Tab. 2) of the surface atomic concentrations is carried out.

| Sample | C% | O% | Au% | Ir% | K% | N% | Ni% | Pd% | Pt% | Rh% | M1/M2 |
|---|---|---|---|---|---|---|---|---|---|---|---|
| AuNi-$CN_l$ 600 | 67.4 | 23.5 | 0.4 | | | | 8.8 | | | | 0.05 |
| IrNi-$CN_l$ 600 | 75.6 | 16.2 | | 0.9 | 1.4 | 3.1 | 2.8 | | | | 0.32 |
| PtIr-$CN_l$ 600 | 82.4 | 12.0 | | 0.6 | 0.9 | 3.6 | | | 0.5 | | 0.80 |
| $PtNi_{1.9}$-$CN_l$ 600 | 75.1 | 14.8 | | | 2.5 | 5.1 | 1.9 | | 0.5 | | 0.26 |
| PtRh-$CN_l$ 600 | 77.4 | 16.7 | | | 1.1 | 3.2 | | | 0.7 | 0.9 | 0.78 |

**Table 2** Surface atomic concentrations for the analysed samples, M1 and M2 are respectively the *"active metal"* and the *"co-catalyst"* as defined at the end of Section 2.2 (unreported concentrations indicate that the elements are below the detection limit of the XPS instrumentation)

The C1s peak and the main peaks for the two M1 and M2 metals in each EC are fitted to identify the number and nature of the chemical environments. The fitting of the C1s peak yields the same results for all the ECs. Accordingly, it will not be further discussed in detail. The main component of the C1s peak always exhibits the lowest energy; it is detected at 284.6 eV and is attributed to the carbon atoms of *"graphitic-like"* carbon nitride sheets where a small concentration of C atoms are substituted by N {Briggs, 1978 #648;Briggs, 1990 #459}{Moulder, 1992 #460}{Wang, 2006 #461}{Di Noto, 2011 #900} Similarly, in all samples the O1s peak displays B.E. values (531.5-532.5 eV) compatible with surface contamination of metal compounds.

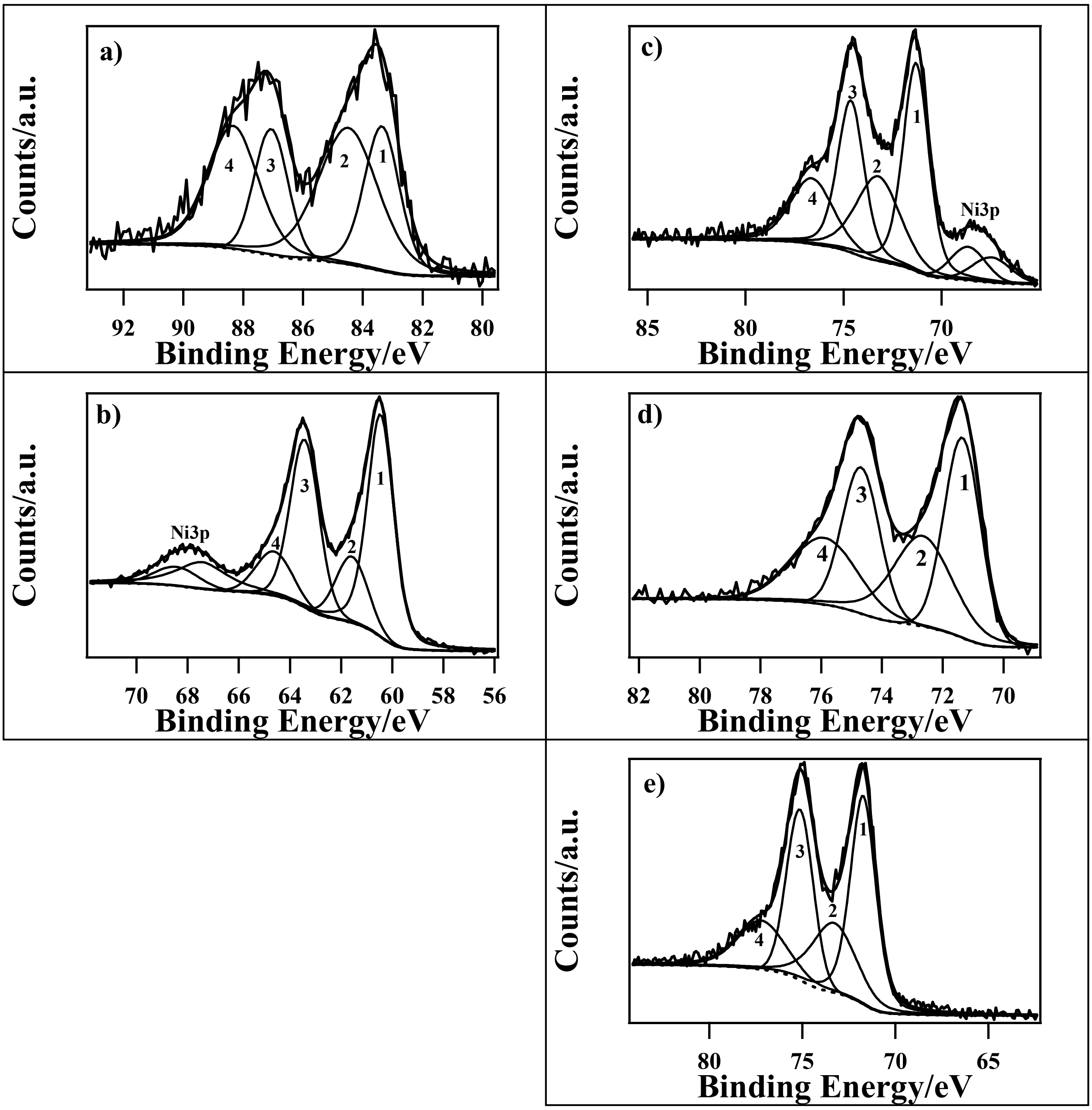

**Figure 1** XPS spectra for the *"active metal"* M1 main region for a) AuNi-$CN_l$ 600 (Au4f), b) IrNi-$CN_l$ 600 (Ir4f), c) $PtNi_{1.9}$-$CN_l$ 600 (Pt4f), d) PtIr-$CN_l$ 600 (Pt4f) and e) PtRh-$CN_l$ 600 (Pt4f). BE values are corrected for surface charging. Peak labels refer to table XX in SI

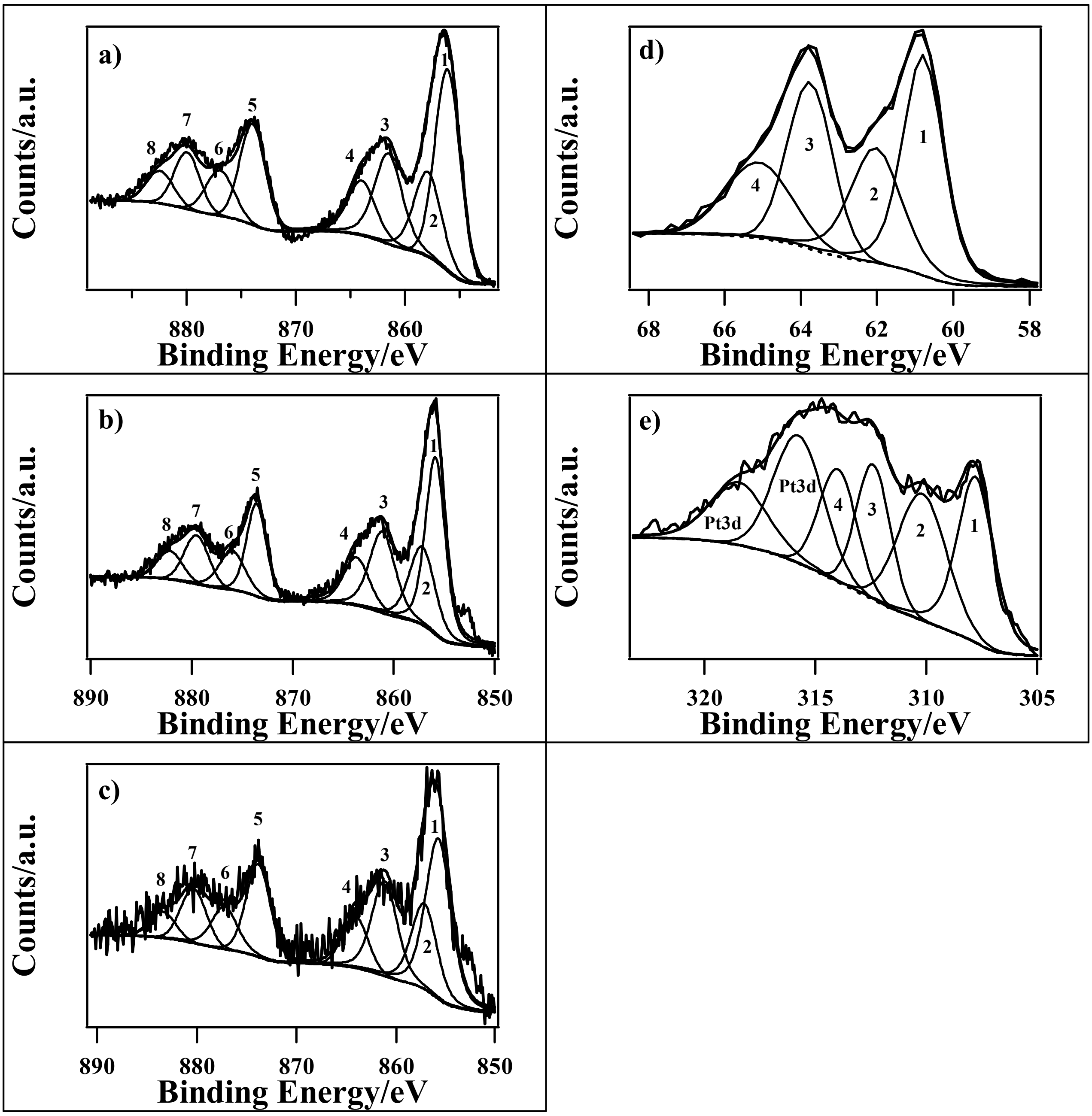

**Figure 2** XPS spectra for the *"co-catalyst"* M2 main region for a) AuNi-$CN_I$ 600 (Ni2p), b) IrNi-$CN_I$ 600 (Ni2p), c) $PtNi_{1.9}$-$CN_I$ 600 (Ni2p), d) PtIr-$CN_I$ 600 (Ir4f) and e) PtRh-$CN_I$ 600 (Rh3d). BE values are corrected for surface charging. Peak labels refer to table XX in SI

General considerations can be made on the XPS results yielded by the *"active metals"* (Fig. 1) and the *"co-catalysts"* (Fig. 2) as follows: as far as the *"active metals"* are concerned, all samples are fitted with two sets of two peaks, with the lower energy component attributed to the species in a metallic state. The higher energy component is assigned depending on the sample: in AuNi-$CN_I$ 600, to possible unidentified $Au^I$ species (possibly partially reacted residue from the precursors, or CN-Au species) {, #458}; in IrNi-$CN_I$ 600 it can be attributed to partially oxidised species (such as residual $IrCl_x$ deriving from the $K_3IrCl_6 \cdot xH_2O$ precursor or CN-Ir species) which several sources report to form in a similar context {, #458}{Moulder, 1992 #460}{Martínez, 2014 #1067}{Tsai, 2012 #1070}{Lin, 2013 #872}{Ziaei-azad, 2013 #876}. Finally, in ECs including Pt as the *"active metal"* (*i.e.* $PtNi_{1.9}$- $CN_I$ 600, PtRh-$CN_I$ 600 and PtIr-$CN_I$ 600 *vide infra*) it is assigned to $Pt^{II}$ {, #458}{Moulder, 1992 #460}{Park, 2002 #899}{Deivaraj, 2003 #1064}, possibly an oxide ($Pt4f_{7/2}$ 72.4 eV), a hydroxide ($Pt4f_{7/2}$ 72.8 eV) or a $Pt^{II}$ specie bonded with the C and N *"coordination nests"* of the CN matrix {Di Noto, 2011 #900}. Concerning the spectra collected from *"co-catalysts"*, the same considerations are made for all the ECs comprising Ni: in all cases two components are identified in the Ni spectra, each consisting of four peaks (from the lowest-energy one to the highest: $Ni2p_{3/2}$, $Ni2p_{3/2\ Sat}$, $Ni2p_{1/2}$ and $Ni2p_{1/2\ Sat}$). It must be noted that all peaks have binding energy values which are by far too high for metallic nickel to be present on the surface (which would generate a $Ni2p_{3/2}$ peak between 852 and 853 eV) {Moulder, 1992 #460}{Briggs, 1978 #648;Briggs, 1990 #459}{, #458}{Hillebrecht, 1983 #1071}; the detected regions (see Figure 2a-2c) are instead characterised by energy values typical for $Ni^{II}$ and $Ni^{III}$ (see Table S3 in SI) {Moulder, 1992 #460}{, #458}. The former region displays peaks which are twice as large as those resulting from the latter region, indicating that there is twice as much $Ni^{II}$ as $Ni^{III}$. This indicates that surface nickel either was never fully reduced or reacted with the environment to form higher oxidation state species, possibly oxohydroxides ($Ni2p_{3/2}$ at 857.3 eV) {, #458}{Moulder, 1992 #460}{Park, 2002 #899}{El Doukkali, 2014 #1065}{Xiong, 2013 #882}. Such species would in part account for the surface oxygen content in the samples. No evidence of NiO is present (binding energy 854.3 eV) {Bharathan, 2014 #867}{, #458}. In the case of ECs PtIr-$CN_I$ 600 and PtRh-$CN_I$ 600, the spectra relative to the *"co-catalysts"* yield results similar to *"active metal"* spectra: two sets of two peaks evidencing a lower energy component relative to a species in a metallic state and a higher energy component attributed respectively to a partially oxidised Ir species (see above) and to $Rh^{III}$, possibly in an oxidic species {, #458}{Moulder, 1992 #460}{Paál, 2007 #887}{Múnera, 2007 #1068}. Tables reporting the binding energies of the collected peaks, as well as those resulting from the fittings can be found in SI (Tab. S2-S3). A more specific examination of each EC is reported below.

### 3.1.1 AuNi-$CN_I$ 600

Several works were published regarding the investigation of Au-Ni systems. Many papers reference Au-Ni alloys, both as nanoparticles {Nishikawa, 2013 #865}{Auten, 2008 #866} and displaying other morphologies (bulk alloys, nanowires etc.) {Lesiak, 2006 #862}{Prasad, 2014 #864}. In all these works however, gold is shown to have an evident tendency to segregate on the surface {Lesiak, 2006 #862} regardless of the Au/Ni ratio employed {Lesiak, 2006 #863}. On the contrary, in our case the Au/Ni surface ratio (0.05) (Tab. 2) is much lower than what was measured through ICP-AES (1.30) (Tab. 1). This could be explained by assuming that the two metals form a Au-Ni *"core-shell"* nanoparticle, with only small traces of gold present on the surface. This is consistent with previous observations performed on similarly prepared catalysts {Di Noto, 2007 #948} which showed that, during synthesis, the VI-period metal (in this case gold) would be the first to undergo reduction nucleating a $M^0$ *"core"*. The IV-period metal (nickel) would form the outer *"shell"* only subsequently. However, records of Au-Ni *"core-shell"* nanonanostructures in the literature are much more sparse {Bharathan, 2014 #867}{Hellenthal, 2012 #868}, especially given a partial bias of the system towards the opposite (*i.e.* Ni-Au) *"core-shell"* structure {Auten, 2008 #866}, and to the best of our knowledge do not feature detailed XPS information.

The measured binding energies in this region are slightly lower than documented in literature for metallic gold, though it must be pointed out that several sources have reported a shift towards lower binding energy values in $AuNi_x$ species due to interaction with a surrounding heterogeneous matrix {Mishra, 2008 #1072} and in particular negative surface charging of the nanoparticle {Jiang, 2007 #1073} by electrodonor species such as the N of the CN matrix involved in *"coordination nests"*.

### 3.1.2 IrNi-$CN_l$ 600

Unlike Au-Ni compounds, very few records are present in the literature directly concerning Ir-Ni systems {Ziaei-Azad, 2014 #869}{Singh, 2010 #870}{Çakanyıldırım, 2012 #871}{Lin, 2013 #872}. Thus, very little specific XPS data is available {Çakanyıldırım, 2012 #871}{Lin, 2013 #872}. Though the calculated surface Ir/Ni ratio (0.32) (Tab. 2) remains lower than the bulk value measured through ICP-AES (0.54) (Tab. 1) the difference is far less significant compared to AuNi-$CN_l$ 600, suggesting that an alloy, rather than a *"core-shell"* system, is obtained. However, the higher nickel content remains consistent with an earlier reduction of the VI-period metal (Ir) followed by the reduction of the IV-period metal (Ni) as observed by Di Noto *et al.* {Di Noto, 2007 #948}. However, due to the aforementioned lack of specific information on the system, no confirmation aside from the obtained experimental data could be provided regarding the tendency of Ni to segregate on the surface relative to Ir. This would however be consistent with the fact that (as with all other prepared M1-Ni compounds) surface nickel is present as $Ni^{II}$ and $Ni^{III}$ rather than $Ni^0$, which suggests a higher tendency to react with the environmental oxygen (Ni is oxophylic) to form species such as oxohydroxides.

Unlike the previous compound, the higher energy component in the *"active metal"* region is much less abundant compared to the lower energy one (peak area ratio approximately 1:3). This indicates that the lower energy component (*i.e.* $Ir^0$) is predominant. Two further peaks are also visible in this region around 68 eV (Fig. 1c): these are however relative to the Ni3p peak {Moulder, 1992 #460}{,  #458} and will not be discussed in further detail.

### 3.1.3 $PtNi_{1.9}$-$CN_l$ 600

Numerous records are present in the literature for this class of bimetallic system{El Doukkali, 2013 #879}{Xiong, 2013 #882}{Mintsouli, 2013 #883}{Wu, 2012 #884}{Cho, 2012 #885}{Çakanyıldırım, 2012 #871}, though several {Navarro, 2014 #880}{Duan, 2013 #881} feature a much lower Pt/Ni ratio (in the order of 5% mol/mol). Even if a large number of these works employ XPS as one of the characterisation methods, an in- depth discussion of the system as synthesised by the method described in this work has not been carried out yet.

As in IrNi-$CN_l$ 600, though the calculated surface Pt/Ni ratio (0.26) (Tab. 2) remains lower than the bulk value measured through ICP-AES (0.52) (Tab. 1) the difference is not as marked as for AuNi-$CN_l$ 600. This suggests that an alloy, rather than a *"core-shell"* system, is obtained, though the high surface Ni content is consistent with observations made for the other ECs.

### 3.1.4 PtIr-$CN_l$ 600

Records of Pt-Ir systems are fairly abundant in the literature, though relatively few of them focus on a bimetallic system and generally deal with oxide systems {da Silva, 2000 #873}, various different Pt-M systems or several heavy metals in general {Ziaei-azad, 2013 #876}{Infantes-Molina, 2007 #875}. However, a few works discussing the PtIr alloy system in nanoparticle form are available, which provide XPS data on the system {Chen, 2011 #874}{Radev, 2012 #877}{Sawy, 2014 #878}.

The Pt/Ir ratio (0.80) measured by XPS (Tab. 2) is quite close to that (0.91) determined by ICP-AES (Tab. 1): this is in agreement with the data found in literature {Sawy, 2014 #878} which reports that surface and bulk metal concentrations in Pt-Ir bimetallic systems are similar. This is also in agreement with the catalyst formation behaviour observed for the Au-Ni and Ir-Ni systems as, in this case, the catalyst consists of two VI-period metals, which would therefore undergo reduction at approximately the same time during the pyrolysis process.

Unlike in the case of region Ir4f (*vide infra*), in the Pt4f region the area ratio between the set of peaks at higher and lower energy is larger (approximately 1:1.2) indicating that nearly equal amounts of $Pt^0$ and $Pt^{II}$ are present. Fitting of the iridium peak yields an ratio between the higher and lower energy peak sets is greater (peak area ratio approximately equal to 1:1.5).

### 3.1.5 PtRh-$CN_l$ 600

This particular bimetallic system presents few records in the literature dedicated to it {Paál, 2007 #887}{Oliveira, 2008 #888}{Kondratenko, 2006 #890}{Park, 2012 #886}{Uemura, 2007 #889} providing in-depth discussion of the oxidation states and chemical environments of the system.

The Pt/Rh atomic ratio (0.78) measured by XPS (Tab. 2) is lower than the value (1.18) evaluated by ICP-AES (Tab. 1). These data indicate a slight tendency of rhodium to segregate on the surface. However, the results are not sufficient to suggest a Pt-Rh *"core-shell"* structure.

The Rh3d region (Fig. 2e) is more difficult to interpret as partial overlap with the $Pt3d_{5/2}$ region {Moulder, 1992 #460} gives rise to a very broad peak. The ratio between surface Rh in the two identified chemical environments (*i.e.* $Rh^0$ and $Rh^{III}$) is *ca.* 1.

### 3.1.6 Investigation of electrochemical performance and interplay between bulk/surface composition

The positive-going voltammetric sweeps of the proposed ECs and of the Pt/C reference in a pure oxygen atmosphere are reported in Figure 3.

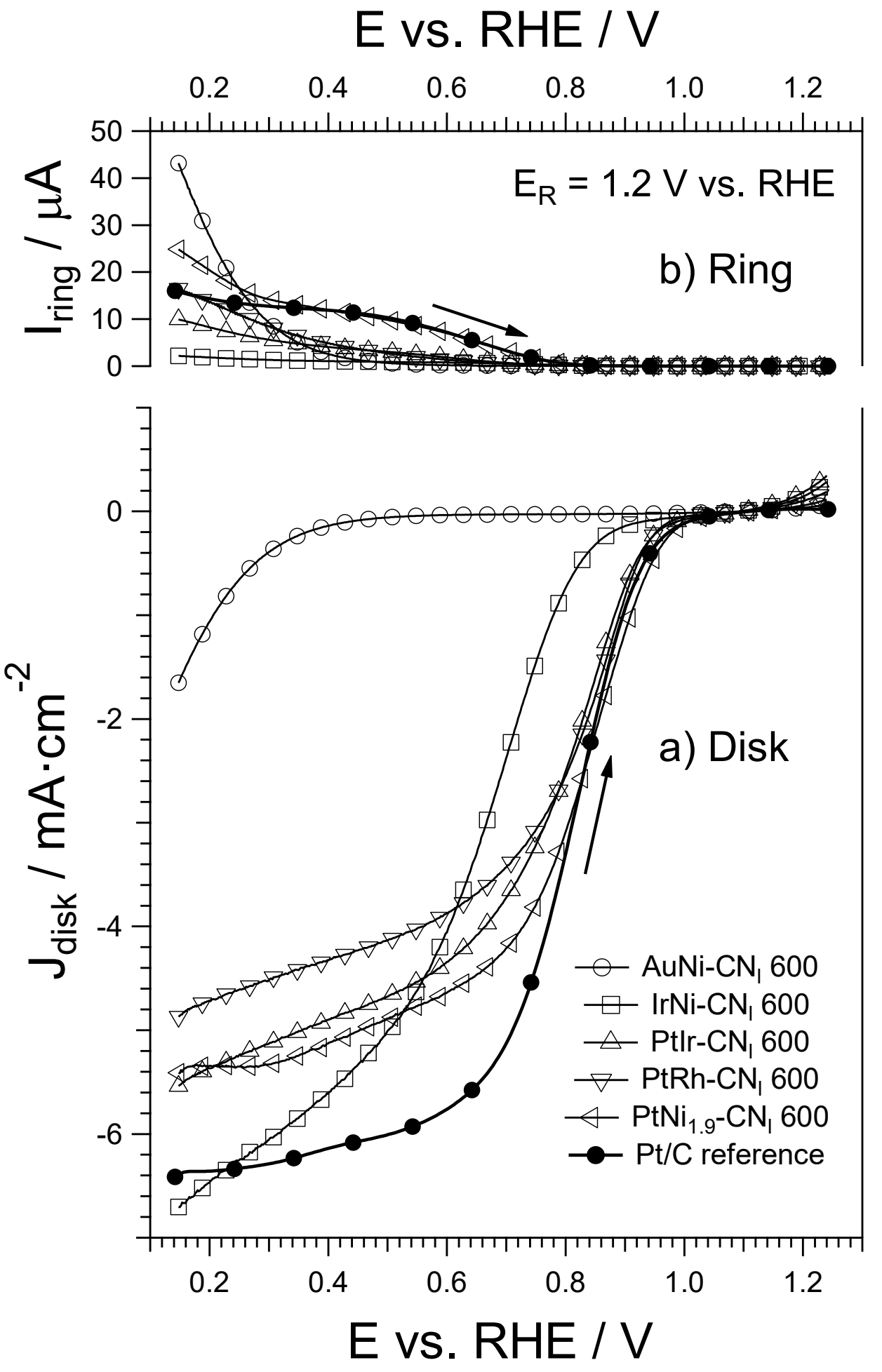

**Figure 3** Positive-going sweeps of the proposed ECs and of the Pt/C reference in a pure oxygen atmosphere. (a) Current densities measured on the glassy-carbon disk; (b) oxidation currents determined on the platinum ring. Cell filled with a 0.1 M $HClO_4$ solution, T = 60 °C, sweep rate = 5 mV·s$^{-1}$, rotation rate 1,600 rpm and $PO_2$ = 1 atm

All the materials exhibit some degree of activity in the ORR. ECs including Pt (*i.e.*, $PtNi_{1.9}$-$CN_l$ 600, PtRh-$CN_l$ 600 and PtIr-$CN_l$ 600) show the lowest ORR overpotentials, which are very similar with one another and with that of the Pt/C reference. The ORR overpotential of IrNi-$CN_l$ 600 is *ca.* 130 mV larger than that of the Pt/C reference. AuNi-$CN_l$ 600 shows the largest ORR overpotential of all the proposed ECs, which is at least 700 mV higher with respect to the Pt/C reference. The ORR current density of AuNi-$CN_l$ 600 is the lowest, to the point that no distinct diffusion-limited plateau can be identified.

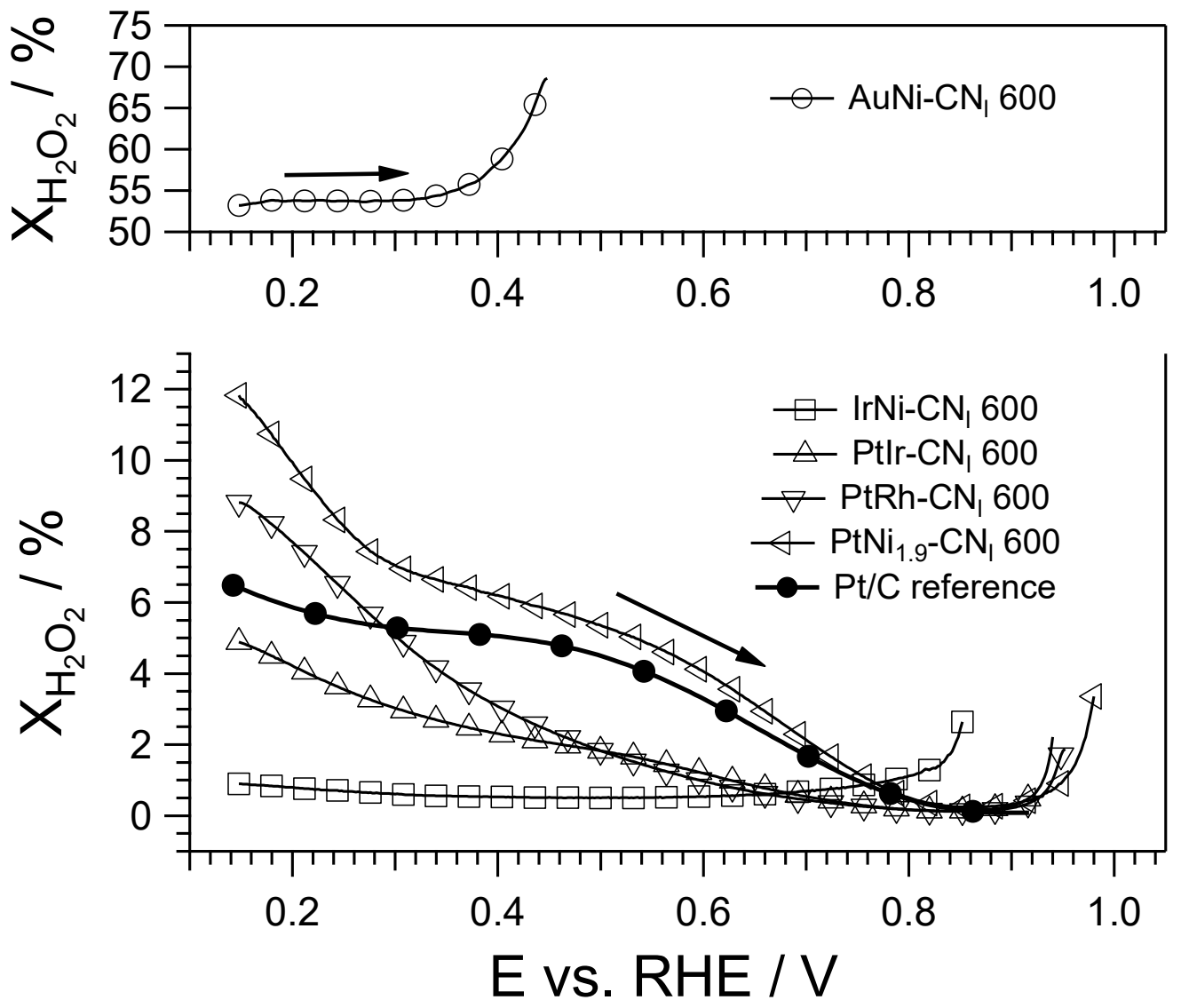

**Figure 4** Fraction of hydrogen peroxide developed during the ORR on the proposed ECs as a function of the potential, determined with Eq. (1) starting from the traces reported in Figure ECHEM1

The fraction of hydrogen peroxide formed during the ORR can be determined on the basis of the disk and ring currents (labeled as $I_D$ and $I_R$, respectively) and the collection efficiency of the platinum ring (N = 0.38), in accordance with Eq. (1) {Vielstich, 2003 #933}{Schmidt, 2001 #932}:

(1)
$$X_{H_2O_2} = \frac{\frac{2I_R}{N}}{I_D + \frac{I_R}{N}}$$

The plot of $XH_2O_2$ as a function of the potential for the proposed ECs and the Pt/C reference is shown in Figure 4. It is possible to study the ORR reaction mechanism at low ORR overpotentials following a procedure which was described elsewhere {Di Noto, 2007 #942}{Di Noto, 2007 #940}{Di Noto, 2010 #939}{Di Noto, 2010 #938}.

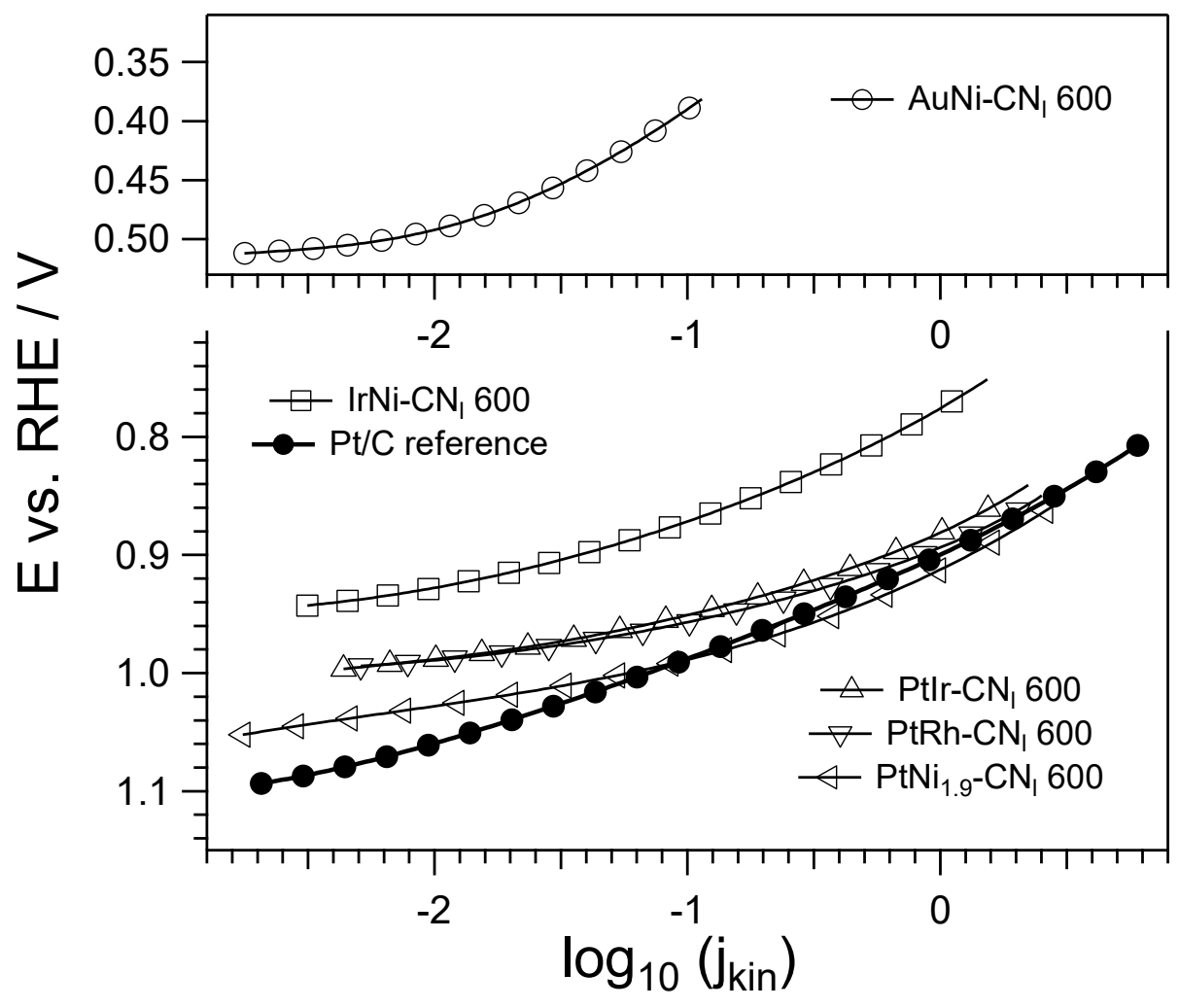

**Figure 5** Tafel plots of the proposed ECs and of the Pt/C reference. Profiles obtained using the traces reported in Figure ECHEM1 after removing the contributions due to mass transport phenomena; currents are normalised on the geometric area of the disk electrode

In detail, the measured current densities are corrected from the contributions due to mass-transport (diffusion) limitations in accordance with Eq. (2):

(2) $$\frac{1}{j} = \frac{1}{j_k} + \frac{1}{j_d} \rightarrow j_k = \frac{j j_d}{j_d - j}$$

In Eq. (2) j, $j_d$ and $j_k$ are the measured current density shown in Figure 3, the diffusion-limited current density and the kinetic current density arising from the electrochemical reaction itself, respectively. $j_d$ is assumed to be the maximum current density in each positive-going sweep in the low-potential region (V ≈ 0.15 V *vs.* RHE). Figure 5 shows the Tafel plots of the ORR for all the investigated materials. With the exception of AuNi-$CN_l$ 600, all the materials show quite similar Tafel plots. In detail, at low ORR overpotentials the Tafel slope is low (*ca.* 40-50 mV·decade$^{-1}$); it increases progressively as the ORR overpotential is raised. This behaviour is typical of the ORR carried out on active sites based on PGMs {Di Noto, 2007 #942}{Di Noto, 2007 #940}{Di Noto, 2010 #939}{Di Noto, 2010 #938}. The overall effectiveness of the ECs is gauged taking into consideration the potentials at $\log_{10}(j_{kin}) = 0$, which are summarised in Table 3.

| Material | Potential (mV *vs.* RHE) |
|---|---|
| *AuNi-$CN_l$ 600* | Lower than 300 mV |
| *IrNi-$CN_l$ 600* | 776 |
| *PtIr-$CN_l$ 600* | 881 |
| *PtRh-$CN_l$ 600* | 894 |
| *$PtNi_{1.9}$-$CN_l$ 600* | 911 |
| *Pt/C reference* | 900 |

**Table 3** Potential of the ECs in the ORR at $\log_{10}(j_{kin}) = 0$.

With respect to the other ECs, the Tafel plot of the ORR carried out by AuNi-$CN_l$ 600 is quite different. At the lower overpotentials, the Tafel slope is very low (*ca.* 20 mV·decade-1) and rises quickly as the overpotential is raised, reaching values larger than 120 mV·decade$^{-1}$. To correlate effectively the electrochemical behaviour of the ECs with their surface chemical composition, the materials are arranged into two classes. The first class, Class I, includes the ECs IrNi-$CN_l$ 600, $PtNi_{1.9}$-$CN_l$ 600 and AuNi-$CN_l$ 600; these materials comprise an *"active metal"* M1 belonging to the sixth period (M1 = Ir, Pt and Au, respectively) together with nickel. The second class, Class II, includes the ECs $PtNi_{1.9}$-$CN_l$ 600, PtRh-$CN_l$ 600 and PtIr-$CN_l$ 600. These ECs comprise platinum together with a *"co-catalyst"* metal M2 belonging to the groups 9 and 10 (M2 = Ni, Rh and Ir, respectively). It is noted that $PtNi_{1.9}$-$CN_l$ 600 belongs to both classes of materials. ICP-AES allows to determine the overall metal content in the materials, averaged on both their surface and the bulk. Thus, the *"average"* molar fraction of the metals M1 and M2 can be calculated for all the materials, taking into account only the transition metal atoms and disregarding all the other elements such as C, N, O, Cl, S and K. In the same way, it is possible to calculate the *"surface"* molar fraction of the metals M1 and M2 starting from the XPS results reported in Table 2.

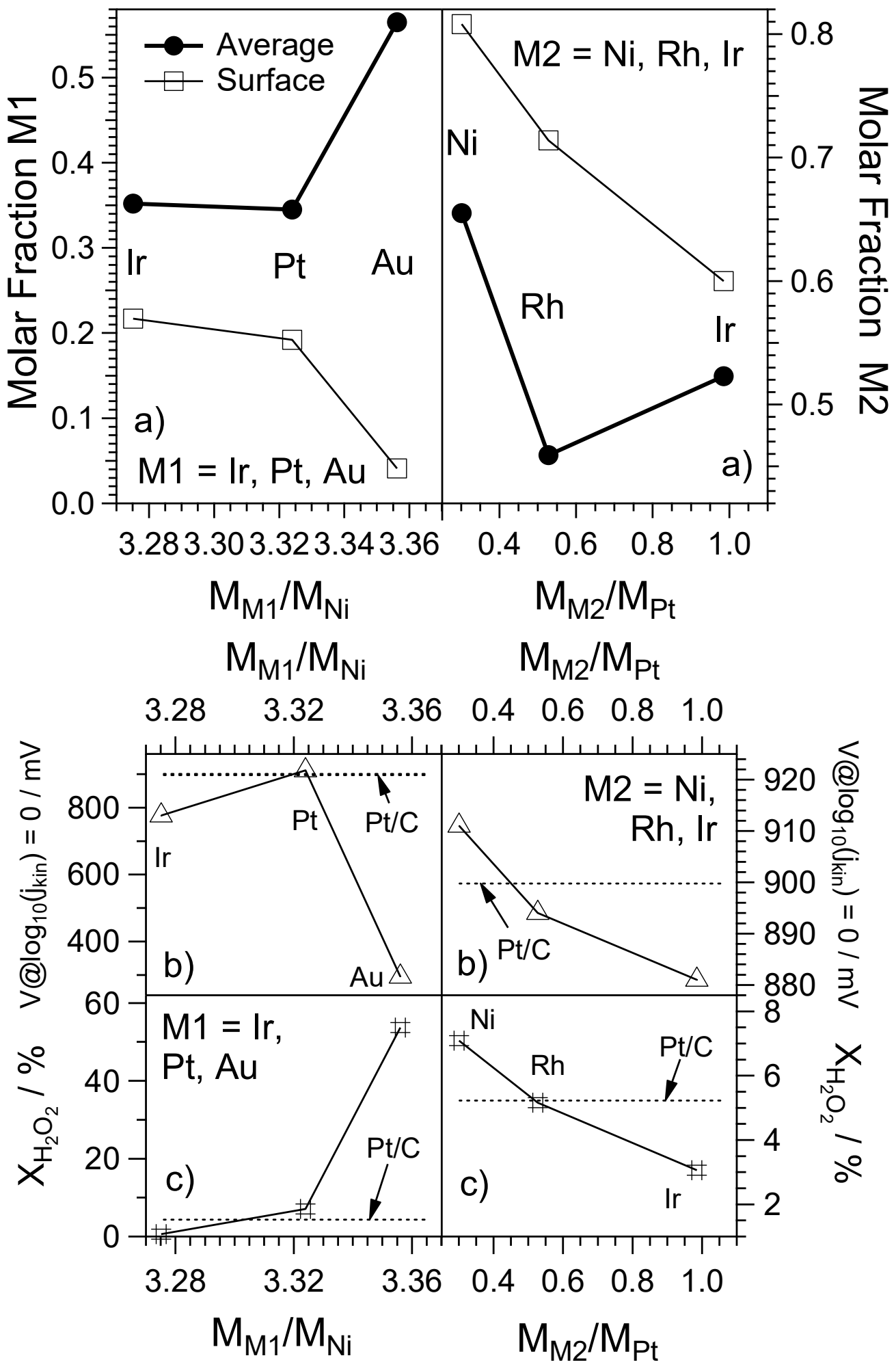

**Figure 6** Correlation between the electrochemical performance of the ECs in the ORR and the surface molar fraction of the metals M1 and M2 in Class I and Class II ECs (left and right panels, respectively). a) Molar fractions of the metals M1 and M2; b) potential of the ECs in the ORR at $log_{10}(j_{kin}) = 0$; c) fraction of hydrogen peroxide in the ORR products at V = 0.3 V *vs.* RHE

Figure 6a compares the *"average"* and the *"surface"* molar fractions of M1 and M2. In the abscissas, the $M_{M1}/M_{Ni}$ and $M_{M2}/M_{Pt}$ ratios are reported, where $M_x$ is the molar mass of the element x, x = M1, Ni, M2 and Pt. It is observed that the *"average"* and the *"surface"* molar fractions of M1 and M2 differ appreciably. In Class I a significant surface segregation of nickel is observed; this effect is particularly marked in the AuNi-$CN_l$ 600 EC. In Class II, the M2 metal also segregates on the surface; this effect is particularly evident for Rh in PtRh-$CN_l$ 600, while it is less important for Ni in $PtNi_{1.9}$-$CN_l$ 600 and even less marked for PtIr-$CN_l$ 600. These evidences are interpreted taking into account the relative ease with which, within a given EC, M1 and M2 undergo reduction during the pyrolysis process, which is the same for all the proposed ECs. It is expected that the tendency of a metal M1 or M2 to undergo reduction and nucleate out of the carbon nitride matrix increases as its oxophilicity is reduced: (i) from Period IV to Period VI; and (ii) from Group 9 to Group 11, following the typical trends observed in transition metals {Basolo, 1967 #1061}. Accordingly, the tendency of the M1 and M2 metals to undergo reduction rises as follows:

Ni ≤ Rh < Ir < Pt < Au. This interpretation allows to explain why AuNi-CN$_l$ 600 exhibits the highest segregation of M2 (*i.e.*, Ni). Indeed, M1 (*i.e.*, Au) undergoes reduction very easily and is nucleated first during the pyrolysis process. M2 is then deposited as a *"shell"* on the Au *"cores"* at the end of the pyrolysis process. On the other hand, in PtIr-CN$_l$ 600 very little segregation of M2 occurs, since Pt and Ir undergo reduction essentially at the same time during the pyrolysis process. One of the most important factors in the determination of the performance of ECs in the ORR is the concentration and reactivity of oxygen-based species adsorbed on the active sites. There are other key parameters affecting these oxygen-based species, such as the pH of the environment and the applied potential. The adsorbed oxygen-based species must be taken into account to correlate the electrochemical performance of the prepared ECs and the information on the surface chemical composition of the latter as determined by XPS. It is well-known that the ORR can be carried out following at least two pathways. In the former, oxygen is reduced directly to water exchanging four electrons (4-electron pathway); in the latter, hydrogen peroxide is produced as an intermediate after the partial reduction of oxygen after the exchange of two electrons (2-electron pathway) {Vielstich, 2003 #933}. A high selectivity in the 4-electron pathway is an important requirement of ORR ECs for application at the cathode of PEM fuel cells. Indeed, hydrogen peroxide is known to degrade the fundamental components of the PEMFC, thus limiting the operating lifetime of the device {Vielstich, 2003 #933}. In an acid environment, if the ORR overpotential is large (*e.g.*, V = 0.3 V *vs.* RHE) it is widely accepted that the 4-electron pathway requires the presence of two neighboring active sites free of adsorbates or other contaminants {Vielstich, 2003 #933}{Schmidt, 2001 #932}. $O_2$ molecules are not expected to adsorb on surface Ni atoms, as the latter are already coordinated by other species (see above). Thus, Ni sites probably act as inhibitors of the 4-electron ORR pathway at large overpotentials. The left panel of Figure 6a shows that the surface molar fraction of Ir and Pt in IrNi-CN$_l$ 600 and PtNi$_{1.9}$-CN$_l$ 600 is quite similar; on the other hand, Figure 6c clearly evidences that at V = 0.3 V *vs.* RHE the selectivity of the former EC in the 4-electron ORR pathway is significantly larger. Indeed, the molar fraction of $H_2O_2$ in the ORR products is *ca.* 1% in IrNi-CN$_l$ 600 *vs.* 7% in PtNi$_{1.9}$-CN$_l$ 600. Thus, it can be concluded that in these conditions the higher selectivity must be ascribed to the Ir atoms which, with respect to Pt atoms, are less clogged by adsorbed oxygen-based species. With respect to all the other ECs presented in this study, AuNi-CN$_l$ 600 is expected to carry out the ORR with a completely different mechanism, which is outlined in the following paragraphs. Accordingly, the selectivity of AuNi-CN$_l$ 600 in the 4-electron mechanism is not discussed further as it is not directly comparable with that of the other ECs. In Class II electrocatalysts, it is observed that the surface molar fraction of the metal M2 decreases in the order: PtNi$_{1.9}$-CN$_l$ 600 > PtRh-CN$_l$ 600 > PtIr-CN$_l$ 600 (see the right panel of Figure 6a. $O_2$ molecules are not expected to adsorb on both $Ni^{II}$/$Ni^{III}$ species and, according to the same rationale, on $Rh^{III}$ species. It is reasonable to assume that $O_2$ molecules can adsorb on $Rh^0$ and $Ir^0$ species, as witnessed by the behaviour of IrNi-CN$_l$ 600. The selectivity of Class II ECs and of the Pt/C reference in the 4-electron ORR pathway increases in the order: PtNi$_{1.9}$-CN$_l$ 600 < Pt/C reference < PtRh-CN$_l$ 600 < PtIr-CN$_l$ 600. As a consequence, it can be observed that this increase in selectivity shows the same trend as the increase in the surface concentration of Pt in Class II ECs (see the right panel of Figure 6c. In addition, since PtRh-CN$_l$ 600 and PtIr-CN$_l$ 600 are actually more selective in the 4-electron ORR pathway with respect to the Pt/C reference, it may be admitted that at V = 0.3 V *vs.* RHE, $Rh^0$ and especially $Ir^0$ species actually help in keeping Pt active sites free of oxygen-based adsorbates, possibly due to an electronic spillover effect.

At low ORR overpotentials the activity of the ECs is bottlenecked by the adsorbed oxygen-based species {Wang, 2009 #931}. As the ORR overpotential decreases, the concentration of adsorbed oxygen-based species increases until all the active sites are blocked. It is expected that Ni atoms are not reduced from the (II) to the (0) oxidation state as the applied potential is raised in an acid environment, as deduced from Pourbaix diagrams {Pourbaix, 1966 #1062} and also considering the binding in C and N *"coordination nests"*. As a consequence, Ni atoms are probably blocked by oxygen-based species and cannot carry out the ORR by themselves. Thus, the main responsible for ORR activity in Class I ECs is expected to be the other metal specie M1, *i.e.* Ir in IrNi-$CN_l$ 600, Pt in $PtNi_{1.9}$-$CN_l$ 600 and Au in AuNi-$CN_l$ 600. With respect to $PtNi_{1.9}$-$CN_l$ 600, IrNi-$CN_l$ 600 becomes inactive in the ORR at an overpotential which is *ca.* 130 mV higher (see the left panel of Figure 6b. Since the surface molar fraction of Ir and Pt are very similar in the two ECs (see the left panel of Figure 6a) it can be admitted that, with respect to Ir, Pt desorbs more easily the ORR intermediates leading to a faster overall ORR kinetic {Vielstich, 2003 #933}. With respect to Pt and Ir, the electrochemical behaviour of Au in the ORR in an acid environment is markedly different. Indeed, Au is essentially inactive in the ORR carried out in an acid environment, as it does not adsorb oxygen readily due to its lack of unpaired d-electrons {Vielstich, 2003 #933}. It is assumed that the small electrochemical activity of AuNi-$CN_l$ 600 in the ORR originates mostly from the CN support matrix. Indeed, carbon-based systems are known to be able to reduce oxygen {Di Noto, 2009 #947}. In particular, if the ORR takes place on the surface of basal graphene-like sheets, a significant production of hydrogen peroxide has been observed {Strelko, 2004 #930}{Strelko, 2000 #929}. This is consistent with the observed results (Fig. 4, upper panel). The results determined on Class I ECs allowed to identify Pt as the most active element in the ORR at low overpotentials. The study of the electrochemical performance of Class II ECs allows to sort out the perturbing effect of a second element on the ORR on platinum active sites. It is observed that, while the surface molar fraction of Pt in Class II ECs increases as follows, $PtNi_{1.9}$-$CN_l$ 600 < PtRh-$CN_l$ 600 < PtIr-$CN_l$ 600, the overall ORR effectiveness of the material shows the inverse trend, as shown in the right panel of Figure 6b. Furthermore, with respect to the Pt/C reference, the potential at $\log_{10}(j_{kin}) = 0$ of $PtNi_{1.9}$-$CN_l$ 600 is higher by ca. 10 mV (see Table 3). This evidence is interpreted as follows. At very low ORR overpotentials, the surface of every Class II EC is almost completely blocked by oxygen-based adsorbed species. The only *"free"* active sites capable to carry out the ORR are located on Pt atoms. Indeed, at low ORR overpotentials Ni and Ir are completely blocked, as justified by the results obtained for Class I ECs. With respect to Ir, Ni is characterised by a smaller atomic radius; thus, the oxygen species adsorbed on Ni are expected to be found nearer to the Pt active sites carrying out the ORR. In addition, Ni is a strong Lewis acid, while Ir is not {Cotton, 1988 #435}. The strongly acid environment created by Ni and its closer proximity to the neighboring Pt active sites promotes the protonation of the intermediates of the ORR. The overall result is a faster ORR kinetics for $PtNi_{1.9}$-$CN_l$ 600, resulting in a lower ORR overpotential with respect to the Ni-free Pt/C reference. Ir does not help to protonate the ORR intermediates as it is not a strong Lewis acid {Cotton, 1988 #435}; on the contrary, Ir probably hinders the desorption of ORR intermediates due to an electronic spillover effect on the neighboring Pt active sites {Vielstich, 2003 #933}. The contribution of Rh at low ORR overpotentials probably falls between the extremes of Ni and Ir. While Rh is not a strong Lewis acid, it is expected to provide a lower electronic spillover effect on Pt active sites with respect to Ir. The overall result is a slight degradation of ORR kinetics, as evidenced in Table 3. In conclusion, in Class II ECs the *"electrochemical"* perturbation effect due to the second element acting as the *"co-catalyst"* is more important than the difference in surface molar fraction of Pt.

The study of the surface concentration ratios between reduced and oxidised species for each metal in the different systems is reported in Table 4. This analysis is carried out on the basis of the proposed fitting of the XPS profiles reported in Figures 1-2, where each metal includes two components.

| | AuNi-$CN_l$ 600 | IrNi-$CN_l$ 600 | PtIr-$CN_l$ 600 | $PtNi_{1.9}$-$CN_l$ 600 | PtRh-$CN_l$ 600 |
|---|---|---|---|---|---|
| $M'_{Red}/M'_{Ox}$ | 0.61[a] | 3.11[c] | <u>1.19</u>[d] | <u>1.44</u>[d] | <u>1.73</u>[d] |
| $M''_{Red}/M''_{Ox}$ | **2.03**[b] | **1.74**[b] | 1.50[c] | **1.88**[b] | 1.09[e] |

[a]$Au^0/Au^I$. [b]$Ni^{II}/Ni^{III}$. [c]$Ir^0/Ir^X$. [d]$Pt^0/Pt^{II}$. [e]$Rh^0/Rh^{III}$

**Table 4** Surface concentration ratios (as evaluated by XPS decomposition) between the reduced and oxidised species for each metal in the ECs.

As can be seen from the first row, in all the examined electrocatalysts with the exception of AuNi-$CN_l$ 600 the VI-period *"active metal"* is found predominantly in a metallic state (*i.e.* oxidation state zero). Thus it can be assumed that, when comparing the ORR performance of the five systems, the 0-state species is the one which is the most involved in the ORR process. In the case of the *"co-catalysts"*, the picture is more complex. In Class II ECs, only one sample includes each of Ni, Rh, and Ir *"co-catalysts"*. Accordingly, it is difficult to sort out with precision the influence of the different oxidation states of each *"co-catalyst"* on the overall ORR performance. On the other hand, Class I ECs only include the Ni *"co-catalyst"*, and it is possible to identify some general trends, as follows. With respect to $Ni^{II}$, $Ni^{III}$ is a stronger Lewis acid {Cotton, 1988 #435}; thus, the latter is expected to play a more important role to promote ORR kinetics. On these bases, it is reasonable to assume that an EC with a higher $Ni^{III}$ content exhibits an improved ORR performance. This is exactly the case for $PtNi_{1.9}$-$CN_l$ 600; this EC, despite having a higher $Ni^{II}/Ni^{III}$ ratio if compared to its IrNi-$CN_l$ 600 analogue, has a higher overall Ni surface concentration (see Table 2) resulting in a higher net $Ni^{III}$ content. In the case of AuNi-$CN_l$ 600, the poor ORR performance does not allow a clear understanding of the role played by the different oxidation states of Ni.

## 4. Conclusions

Five bimetallic carbon nitride-supported ECs are synthesised. The ECs are divided into two classes, Class I and Class II. Class I is made up by bimetallic ECs including both a VI-period metal (*i.e.*, Ir, Pt, and Au) and Ni. Class II comprises materials including both Pt and either Ni, Rh and Ir. One EC (*i.e.*, $PtNi_{1.9}$-$CN_l$ 600) belongs to both Class I and Class II. The bulk chemical composition of the ECs is determined by ICP-AES and microanalysis, while detailed insight on the surface composition is achieved by high-resolution XPS analyses. The ORR performance and reaction mechanism of the ECs is determined by CV-TF-RRDE tests. The results allow to: (i) study the distribution of the metallic elements in the active sites; and (ii) correlate the surface composition of the ECs with the electrochemical performance, identifying the roles played by the various metal species. At high ORR overpotentials (*e.g.*, at *ca.* 0.3 V *vs.* RHE), the selectivity in the 4-electron mechanism is promoted by Ir and Rh species. The latter help reduce the surface concentration of O-based adsorbates, which hinder the dissociative adsorption of $O_2$ yielding water as the final reaction product. It is demonstrated that the ORR performance of the ECs is mostly dependent on the VI-period element (*i.e.*, Ir, Pt, Au), which plays the role of *"active metal"*. The best turnover frequency is achieved for ECs including Pt; indeed, this latter element strikes the best compromise between a facile $O_2$ adsorption and an easy transfer of electrons to form reduced product species (*i.e.*, $H_2O$ and $H_2O_2$). The second metal (Ni, Rh, Ir) acts as a *"co-catalyst"* for the more highly performing *"active metal"*; it is shown that this *"co-catalyst"* is able to promote the kinetics of the ORR by facilitating the protonation of the intermediates and their final desorption from the surface of the EC. This effect is maximised as the Lewis acid character of the co-catalyst is raised (Ir < Rh < Ni). It is further shown that the VI-period metal is mostly in its (0) oxidation state, which is expected to play the most significant role in the ORR process. On the other hand, two oxidation states are always detected on the *"co-catalyst"*; the highest (*i.e.*, $Ir^{x}$, $Rh^{III}$ and $Ni^{III}$) is the one which is most active to enhance the ORR turnover frequency. The metal active sites of the ECs are stabilised in C and N *"coordination nests"* of the CN matrix. Taken together, the best results are registered for $PtNi_{1.9}$-$CN_{l}$ 600, which comprises both the best *"active metal"* (*i.e.*, Pt in its (0) oxidation state) and the most effective *"co-catalyst"* (*i.e.*, $Ni^{III}$ with its strong Lewis acid character). Indeed, $PtNi_{1.9}$-$CN_{l}$ 600 exhibits an improved ORR performance with respect to the Pt/C reference, and demonstrates the relevance of these studies in the ongoing quest for the development of improved ORR electrocatalysts with a high turnover frequency for application in next-generation PEMFCs.

## 5. Acknowledgements

This work was funded by the Strategic Project of the University of Padova "From Materials for Membrane-Electrode Assemblies to Energy Conversion and Storage Devices - MAESTRA". The research leading to these results has received funding from the European Union Seventh Framework Programme under grant agreement n°604391 Graphene Flagship. The research leading to these results has received funding from the European Horizon 2020 under grant agreement n°696656 Graphene Flagship.

## 6. Notes and references

## Supplementary information

| Material | Solution A | Solution B | Solution C | A/B molar ratio |
|---|---|---|---|---|
| AuNi-CN$_I$ 600 | 1.3318 g $KAuCl_4$ | 1.0333 g $K_2Ni(CN)_4{\cdot}xH_2O$ | 4.70 g (D+)-sucrose | 1:1 |
| IrNi-CN$_I$ 600 | 1.000 g $K_3IrCl_6{\cdot}xH_2O$ | 0.6619 $K_2Ni(CN)_4{\cdot}xH_2O$ | 3.00 g (D+)-sucrose | 1:1 |
| PtIr-CN$_I$ 600 | 1.000 g $K_3IrCl_6{\cdot}xH_2O$ | 0.823 g $K_2Pt(CN)_4{\cdot}xH_2O$ | 4.27 g (D+)-sucrose | 1:1 |
| PtNi$_{1.9}$-CN$_I$ 600 | 0.832 g $K_2PtCl_4$ | 1.288 g $K_2Ni(CN)_4{\cdot}xH_2O$ | 4.46 g (D+)-sucrose | 1:1.9 |
| PtRh-CN$_I$ 600 | 0.540 g $RhCl_3{\cdot}xH_2O$ | 0.861 g $K_2Pt(CN)_4{\cdot}xH_2O$ | 4.46 g (D+)-sucrose | 1:1 |

**Table S1** Labels used to identify the ECs presented in this paper; amounts of reagents used in their preparation; and nominal molar ratios between the A and B metals.

| EC | O1s (eV) | Au4f$_{7/2}$ (eV) | Au4f$_{5/2}$ (eV) | Ni2p$_{3/2}$ (eV) | Ni2p$_{1/2}$ (eV) | Ni $L_2M_{23}M_{45}$ ($^1P$) Auger Parameter (eV) |
|---|---|---|---|---|---|---|
| AuNi-CN$_I$ 600 | 531.4 | 83.6 | 87.5 | 856.4 | 873.8 | 1697.6 |
| | O1s (eV) | Ir4f$_{7/2}$ (eV) | Ir4f$_{5/2}$ (eV) | Ni2p$_{3/2}$ (eV) | Ni2p$_{1/2}$ (eV) | Ni $L_2M_{23}M_{45}$ ($^1P$) Auger Parameter (eV) |
| IrNi-CN$_I$ 600 | 531.4 | 60.5 | 63.5 | 855.8 | 873.9 | 1698.0 |
| | O1s (eV) | Pt4f$_{7/2}$ (eV) | Pt4f$_{5/2}$ (eV) | Ni2p$_{3/2}$ (eV) | Ni2p$_{1/2}$ (eV) | N/A |
| PtNi$_{1.9}$-CN$_I$ 600 | 531.8 | 71.3 | 74.5 | 856.1 | 873.8 | N/A |
| | O1s (eV) | Pt4f$_{7/2}$ (eV) | Pt4f$_{5/2}$ (eV) | Ir4f$_{7/2}$ (eV) | Ir4f$_{5/2}$ (eV) | / |
| PtIr-CN$_I$ 600 | 532.6 | 71.4 | 73.7 | 61.0 | 63.8 | / |
| | O1s (eV) | Pt4f$_{7/2}$ (eV) | Pt4f$_{5/2}$ (eV) | Rh3d$_{5/2}$ (eV) | Rh3d$_{3/2}$ (eV) | / |
| PtRh-CN$_I$ 600 | 532.0 | 71.9 | 74.9 | 307.6 | 312.6 | / |

**Table S2** Corrected binding energies and Auger Parameter for the main peaks in sample AuNi-CN$_I$ 600

| EC | Peak component | Corrected binding energy (eV) | Peak Number | Assignment |
|---|---|---|---|---|
| AuNi-CN$_I$ 600 | Au4f$_{7/2}$ | 83.4 | 1 | $Au^0$ |
| AuNi-CN$_I$ 600 | Au4f$_{5/2}$ | 87.1 | 3 | $Au^0$ |
| AuNi-CN$_I$ 600 | Au4f$_{7/2}$ | 84.5 | 2 | $Au^I$ |
| AuNi-CN$_I$ 600 | Au4f$_{5/2}$ | 88.3 | 4 | $Au^I$ |
| AuNi-CN$_I$ 600 | Ni2p$_{3/2}$ | 856.1 | 1 | $Ni^{II}$ |
| AuNi-CN$_I$ 600 | Ni2p$_{3/2\ Sat}$ | 861.5 | 3 | $Ni^{II}$ |
| AuNi-CN$_I$ 600 | Ni2p$_{1/2}$ | 874.0 | 5 | $Ni^{II}$ |
| AuNi-CN$_I$ 600 | Ni2p$_{1/2\ Sat}$ | 880.0 | 7 | $Ni^{II}$ |
| AuNi-$CN_I$ 600 | Ni2$p_{3/2}$ | 857.9 | 2 | $Ni^{III}$ |
| AuNi-$CN_I$ 600 | Ni2$p_{3/2\ Sat}$ | 864.0 | 4 | $Ni^{III}$ |
| AuNi-$CN_I$ 600 | Ni2$p_{1/2}$ | 877.0 | 6 | $Ni^{III}$ |
| AuNi-$CN_I$ 600 | Ni2$p_{1/2\ Sat}$ | 882.5 | 8 | $Ni^{III}$ |
| IrNi-$CN_I$ 600 | Ir4$f_{7/2}$ | 60.5 | 1 | $Ir^0$ |
| IrNi-$CN_I$ 600 | Ir4$f_{5/2}$ | 63.4 | 3 | $Ir^0$ |
| IrNi-$CN_I$ 600 | Ir4$f_{7/2}$ | 61.6 | 2 | $Ir^x$ |
| IrNi-$CN_I$ 600 | Ir4$f_{5/2}$ | 64.6 | 4 | $Ir^x$ |
| IrNi-$CN_I$ 600 | Ni2$p_{3/2}$ | 855.9 | 1 | $Ni^{II}$ |
| IrNi-$CN_I$ 600 | Ni2$p_{3/2\ Sat}$ | 861.1 | 3 | $Ni^{II}$ |
| IrNi-$CN_I$ 600 | Ni2$p_{1/2}$ | 873.6 | 5 | $Ni^{II}$ |
| IrNi-$CN_I$ 600 | Ni2$p_{1/2\ Sat}$ | 879.5 | 7 | $Ni^{II}$ |
| IrNi-$CN_I$ 600 | Ni2$p_{3/2}$ | 857.2 | 2 | $Ni^{III}$ |
| IrNi-$CN_I$ 600 | Ni2$p_{3/2\ Sat}$ | 863.7 | 4 | $Ni^{III}$ |
| IrNi-$CN_I$ 600 | Ni2$p_{1/2}$ | 876.0 | 6 | $Ni^{III}$ |
| IrNi-$CN_I$ 600 | Ni2$p_{1/2\ Sat}$ | 882.1 | 8 | $Ni^{III}$ |
| $PtNi_{1.9}$-$CN_I$ 600 | Pt4$f_{7/2}$ | 71.3 | 1 | $Pt^0$ |
| $PtNi_{1.9}$-$CN_I$ 600 | Pt4$f_{5/2}$ | 74.6 | 3 | $Pt^0$ |
| $PtNi_{1.9}$-$CN_I$ 600 | Pt4$f_{7/2}$ | 73.2 | 2 | $Pt^{II}$ |
| $PtNi_{1.9}$-$CN_I$ 600 | Pt4$f_{5/2}$ | 76.6 | 4 | $Pt^{II}$ |
| $PtNi_{1.9}$-$CN_I$ 600 | Ni2$p_{3/2}$ | 855.7 | 1 | $Ni^{II}$ |
| $PtNi_{1.9}$-$CN_I$ 600 | Ni2$p_{3/2\ Sat}$ | 861.1 | 3 | $Ni^{II}$ |
| $PtNi_{1.9}$-$CN_I$ 600 | Ni2$p_{1/2}$ | 873.8 | 5 | $Ni^{II}$ |
| $PtNi_{1.9}$-$CN_I$ 600 | Ni2$p_{1/2\ Sat}$ | 880.3 | 7 | $Ni^{II}$ |
| $PtNi_{1.9}$-$CN_I$ 600 | Ni2$p_{3/2}$ | 857.2 | 2 | $Ni^{III}$ |
| $PtNi_{1.9}$-$CN_I$ 600 | Ni2$p_{3/2\ Sat}$ | 864.1 | 4 | $Ni^{III}$ |
| $PtNi_{1.9}$-$CN_I$ 600 | Ni2$p_{1/2}$ | 877.2 | 6 | $Ni^{III}$ |
| $PtNi_{1.9}$-$CN_I$ 600 | Ni2$p_{1/2\ Sat}$ | 883.4 | 8 | $Ni^{III}$ |
| PtIr-$CN_I$ 600 | Pt4$f_{7/2}$ | 71.4 | 1 | $Pt^0$ |
| PtIr-$CN_I$ 600 | Pt4$f_{5/2}$ | 74.7 | 3 | $Pt^0$ |
| PtIr-$CN_I$ 600 | Pt4$f_{7/2}$ | 72.7 | 2 | $Pt^{II}$ |
| PtIr-$CN_I$ 600 | Pt4$f_{5/2}$ | 75.9 | 4 | $Pt^{II}$ |
| PtIr-$CN_I$ 600 | Ir4$f_{7/2}$ | 60.8 | 1 | $Ir^0$ |
| PtIr-$CN_I$ 600 | Ir4$f_{5/2}$ | 63.8 | 3 | $Ir^0$ |
| PtIr-$CN_I$ 600 | $Ir4f_{7/2}$ | 62.0 | 2 | $Ir^x$ |
| PtIr-$CN_I$ 600 | $Ir4f_{5/2}$ | 65.0 | 4 | $Ir^x$ |
| PtRh-$CN_I$ 600 | $Pt4f_{7/2}$ | 71.7 | 1 | $Pt^0$ |
| PtRh-$CN_I$ 600 | $Pt4f_{5/2}$ | 75.1 | 3 | $Pt^0$ |
| PtRh-$CN_I$ 600 | $Pt4f_{7/2}$ | 73.3 | 2 | $Pt^{II}$ |
| PtRh-$CN_I$ 600 | $Pt4f_{5/2}$ | 77.2 | 4 | $Pt^{II}$ |
| PtRh-$CN_I$ 600 | $Rh3d_{5/2}$ | 307.8 | 1 | $Rh^0$ |
| PtRh-$CN_I$ 600 | $Rh3d_{3/2}$ | 312.4 | 3 | $Rh^0$ |
| PtRh-$CN_I$ 600 | $Rh3d_{5/2}$ | 310.2 | 2 | $Rh^{III}$ |
| PtRh-$CN_I$ 600 | $Rh3d_{3/2}$ | 314.0 | 4 | $Rh^{III}$ |

**Table S3** XPS decomposition components and their assignments